\documentclass[a4paper]{article} 

\usepackage[numbers,square,sort]{natbib}
\usepackage{graphicx}
\usepackage[usenames,dvipsnames]{color}
\usepackage{amssymb,amsmath,amsthm}
\usepackage{subfigure}
\usepackage[subfigure]{tocloft}
\usepackage{framed}
\usepackage{hyperref}
\hypersetup{pdfborder = {0 0 0}} 
\usepackage{listings}
\usepackage{bibunits}

\lstdefinelanguage{lua}
  {morekeywords={and,break,do,else,elseif,end,false,for,function,if,in,local,
     nil,not,or,repeat,return,then,true,until,while},
   morekeywords={[2]arg,assert,collectgarbage,dofile,error,_G,getfenv,
     getmetatable,ipairs,load,loadfile,loadstring,next,pairs,pcall,print,
     rawequal,rawget,rawset,select,setfenv,setmetatable,tonumber,tostring,
     type,unpack,_VERSION,xpcall},
   morekeywords={[2]coroutine.create,coroutine.resume,coroutine.running,
     coroutine.status,coroutine.wrap,coroutine.yield},
   morekeywords={[2]module,require,package.cpath,package.load,package.loaded,
     package.loaders,package.loadlib,package.path,package.preload,
     package.seeall},
   morekeywords={[2]string.byte,string.char,string.dump,string.find,
     string.format,string.gmatch,string.gsub,string.len,string.lower,
     string.match,string.rep,string.reverse,string.sub,string.upper},
   morekeywords={[2]table.concat,table.insert,table.maxn,table.remove,
   table.sort},
   morekeywords={[2]math.abs,math.acos,math.asin,math.atan,math.atan2,
     math.ceil,math.cos,math.cosh,math.deg,math.exp,math.floor,math.fmod,
     math.frexp,math.huge,math.ldexp,math.log,math.log10,math.max,math.min,
     math.modf,math.pi,math.pow,math.rad,math.random,math.randomseed,math.sin,
     math.sinh,math.sqrt,math.tan,math.tanh},
   morekeywords={[2]io.close,io.flush,io.input,io.lines,io.open,io.output,
     io.popen,io.read,io.tmpfile,io.type,io.write,file:close,file:flush,
     file:lines,file:read,file:seek,file:setvbuf,file:write},
   morekeywords={[2]os.clock,os.date,os.difftime,os.execute,os.exit,os.getenv,
     os.remove,os.rename,os.setlocale,os.time,os.tmpname},     
   morekeywords={[2]sphere,fgu,add,meshnode,vertexlist,vec3,sin,distance,iso,inset,capov,flattenvl,new_lsystem},   
   sensitive=true,
   morecomment=[l]{--},
   morecomment=[s]{--[[}{]]--},
   morestring=[b]",
   morestring=[d]'
  }
\definecolor{lightgray}{gray}{0.5}
\newcommand{\func}[1]{\noindent \texttt{#1}} 
\newcommand{\code}[1]{\noindent \texttt{#1}} 

\graphicspath{{Figures/}}

\title{A scriptable, generative modelling system for dynamic 3D meshes}
\author{Jon McCormack\\
Faculty of Information Technology, Monash University\\
Caulfield East 3145, Victoria, Australia\\
\texttt{Jon.McCormack@monash.edu}\\
 \and
  Ben Porter \\
  \texttt{benjamin.porter@gmail.com}\\
  \and
  James Wetter\\
  \texttt{wetter.j@gmail.com}}

\date{1 May 2012}

\begin{document}
\maketitle

\begin{abstract}
We describe a flexible, script-based system for the procedural generation and animation of 3D geometry. Dynamic triangular meshes are generated through the real-time execution of scripts written in the Lua programming language. Tight integration between the programming environment, runtime engine and graphics visualisation enables a workflow between coding and visual results that encourages experimentation and rapid prototyping. The system has been used successfully to generate a variety of complex, dynamic organic forms including complex branching structures, scalable symmetric manifolds and abstract organic forms. We use examples in each of these areas to detail the main features of the system, which include a set of flexible 3D mesh operations integrated with a Lua-based L-system interpreter that creates geometry using generalised cylinders.
\end{abstract}

\section{Introduction}

\begin{figure}
  \centering
    \subfigure[]{
    \includegraphics[width=.3\textwidth]{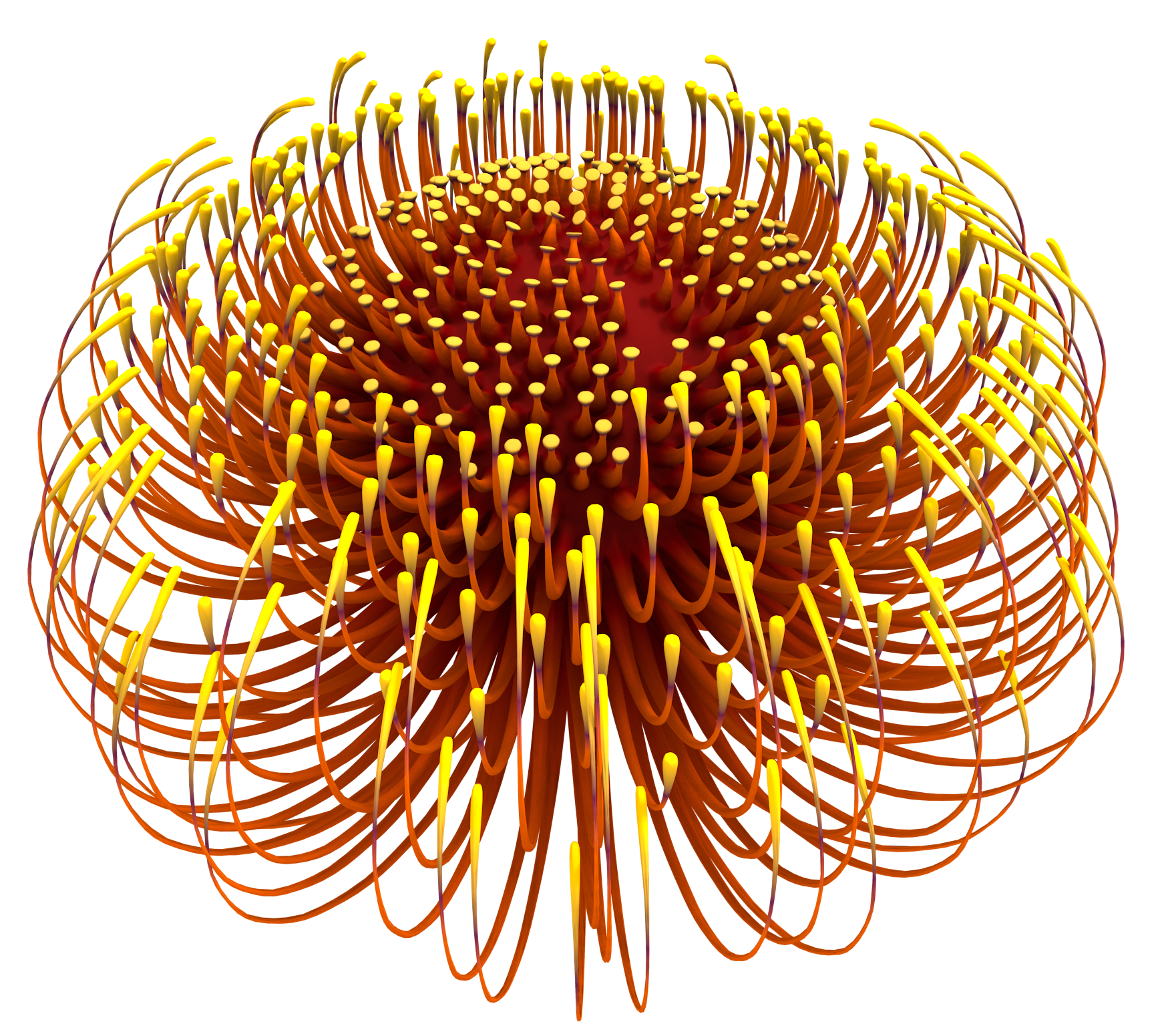}
    \label{fig:intro:flower}
    }
    \subfigure[]{
    \includegraphics[width=.3\textwidth]{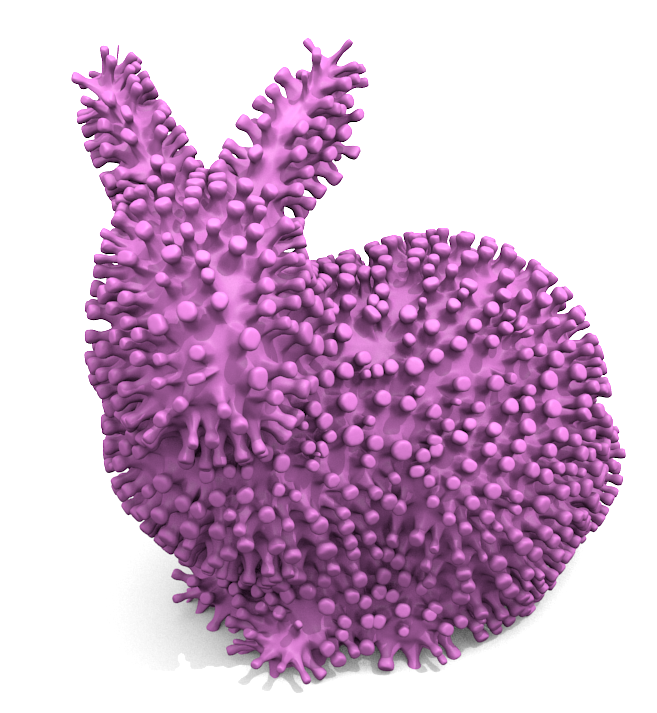}
    \label{fig:intro:bunny}
    }
    \subfigure[]{
    \includegraphics[width=.3\textwidth]{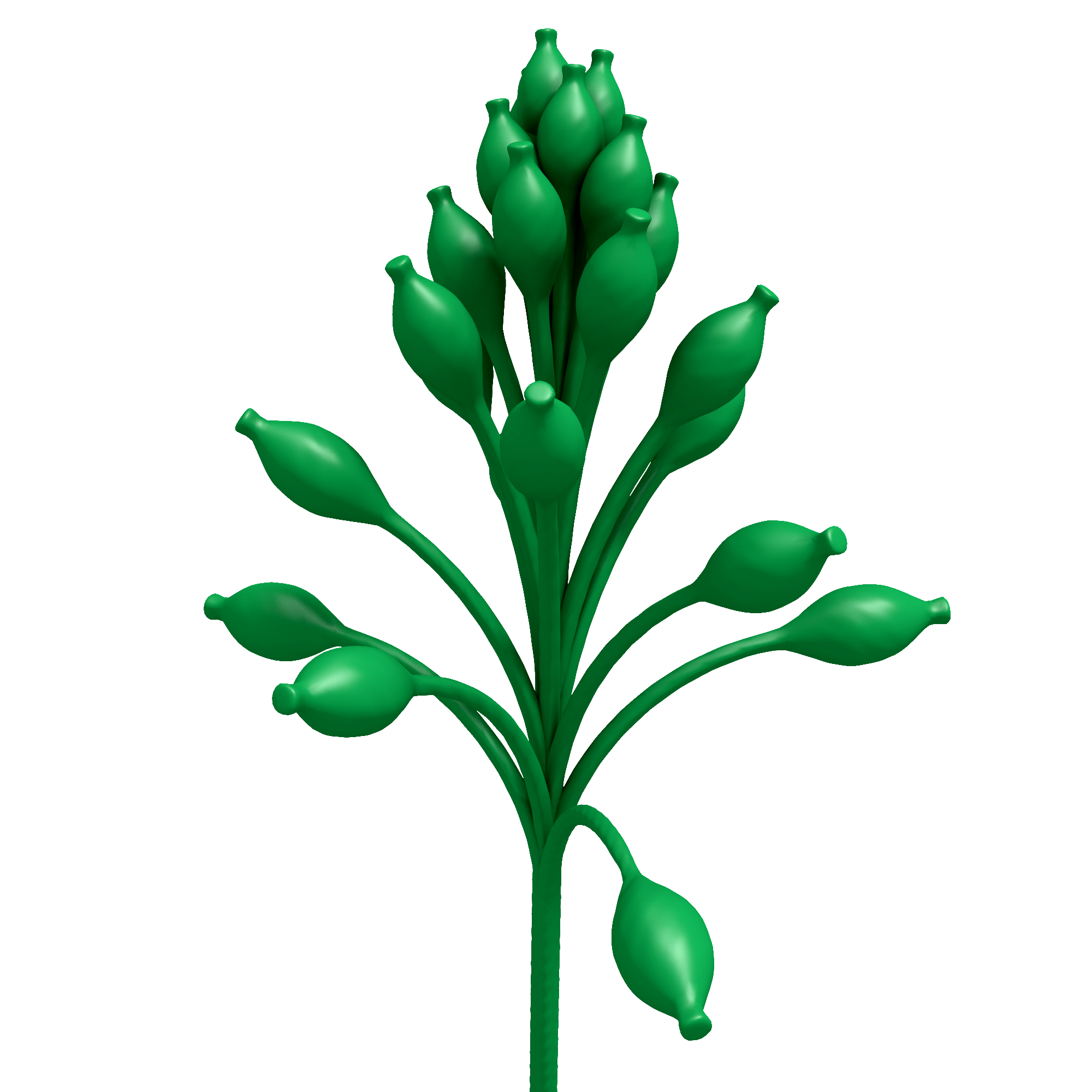}
    \label{fig:intro:plant}
    }
    \label{fig:intro}
  \caption{Example models created with the software described in this paper. (a) A Waratah (\emph{Telopea}) flower created with the extrusion script described in Section \ref{ss:example:extrusion}. (b) A stalk-covered rabbit generated by applying the extrusion script to the Stanford Bunny model. (c) A \emph{Calcispongiae} design modelled with timed L-systems, generalised cylinders, and Lua scripting.}
\end{figure}

Modelling of complex, organic forms presents an enormous challenge in an architectural and design context. One successful approach has been the use of procedural modelling, where the designer supplies a procedure that the computer executes to build and animate a geometric model \cite{Ebert2003}. Typically the designer specifies form and behaviour at a high-level and the computer ``fills in the details'', generating the necessary geometric, behavioural and surface complexity automatically. Popular procedural techniques that generate complex organic structures include: L-systems \cite{Prusinkiewicz1996}, CDM \cite{McCormack2005}, Cellular models  \cite{Fleischer1995b}, Physical Models \cite{Combaz2006}, Vertex-Vertex Systems \cite{Smith2006}, and the Simplicial Developmental System \cite{Porter2010}, to name just a few.

\subsection{Script-based Modelling and Development} 
The recent proliferation of code-based creative systems suggests that programming is now increasingly embraced by architects and designers as a creative medium \cite{Reas2007,openFrameworks,Smith2008}. These systems, ideal for procedural modelling, have enabled many new and creatively interesting results, e.g.\ \cite{Reas2010}. They are successful because they combine the flexibility of programming with a specialised palette of custom functions and support libraries. They hide the underlying complexity of library functionality from the user, presenting a simple, but powerful interface that makes design and experimentation easy.

Professional animation packages support scripting languages too, using either a custom language (e.g. Maya's \emph{MEL}, Cinema 4D's \emph{COFFEE}) or a general purpose language, such as Python or Java. General purpose languages have the advantage of a broad user community, extensive use and testing, good documentation, examples, and potentially, user familiarity. Irrespective of their origins however, when employed in 3D design systems these languages must balance programming functionality with the interactive modelling and animation components of the software, where scripting is just one feature of many. So while they typically support the creation and manipulation of 3D form, a procedural approach is not necessarily their main focus. For example, the scripting of meshes with changing topologies is not readily supported in a number of systems. This makes the flow of ideas between script and dynamic form less effective than it could be.

This paper describes \emph{Fugu}, a script-driven 3D modelling system developed by the authors. Fugu permits flexible, procedural modelling of dynamic geometric meshes. It supports the generation, manipulation and animation of 3D form using scripts written in the programming language \emph{Lua} \cite{Ierusalimschy2006}, with the specific goal of easily creating the complex organic features found in natural forms and other complex geometries. The system is simple: a modelling program is a single Lua module, and the user interface provides the means to program and execute the code, then visualise and interact with the resulting model cohesively.

We will illustrate Fugu's approach to modelling using examples that highlight different aspects of its functionality and features. Fugu is designed to support multiple levels of model representation and control using the modularity provided by Lua. The goal is to provide a tool that allows easy definition and manipulation of a multi-representation model (for example, mesh and armature representations) within a single script.

After describing the operational and interface basics of Fugu in the next section, we will illustrate different aspects of its functionality in Sections \ref{ss:example:extrusion} and \ref{ss:example:lsystem}, before discussing the effectiveness of our approach and looking at how the system could be further developed in Section \ref{s:conclusions}.

\section{Fugu}

The basis of modelling in Fugu is a Lua script that generates and manipulates 3D triangular meshes via discrete simulation. Lua is a powerful, efficient, lightweight scripting language, frequently used in computationally demanding real-time applications such as games and multimedia environments \cite{Smith2008}. We chose Lua because of its simple syntax, extensible semantics and the ease with which it can be embedded in other software.

Lua scripts interface to a dynamic 3D runtime engine, written in C++, which handles the generation and display of geometry, along with the Fugu user interface (Section \ref{ss:fugu:interface}). Fugu scripts are single file Lua modules containing module-scoped variables and two special callback functions, \func{setup()} and \func{update()} (Section \ref{ss:fugu:script}). An array of functions and libraries (Section \ref{ss:fugu:functionality}) are exposed to the scripting system to manipulate and control this runtime engine.

\subsection{Interface}\label{ss:fugu:interface}

\begin{figure}[htbp]
    \centering
    \includegraphics[width=\textwidth]{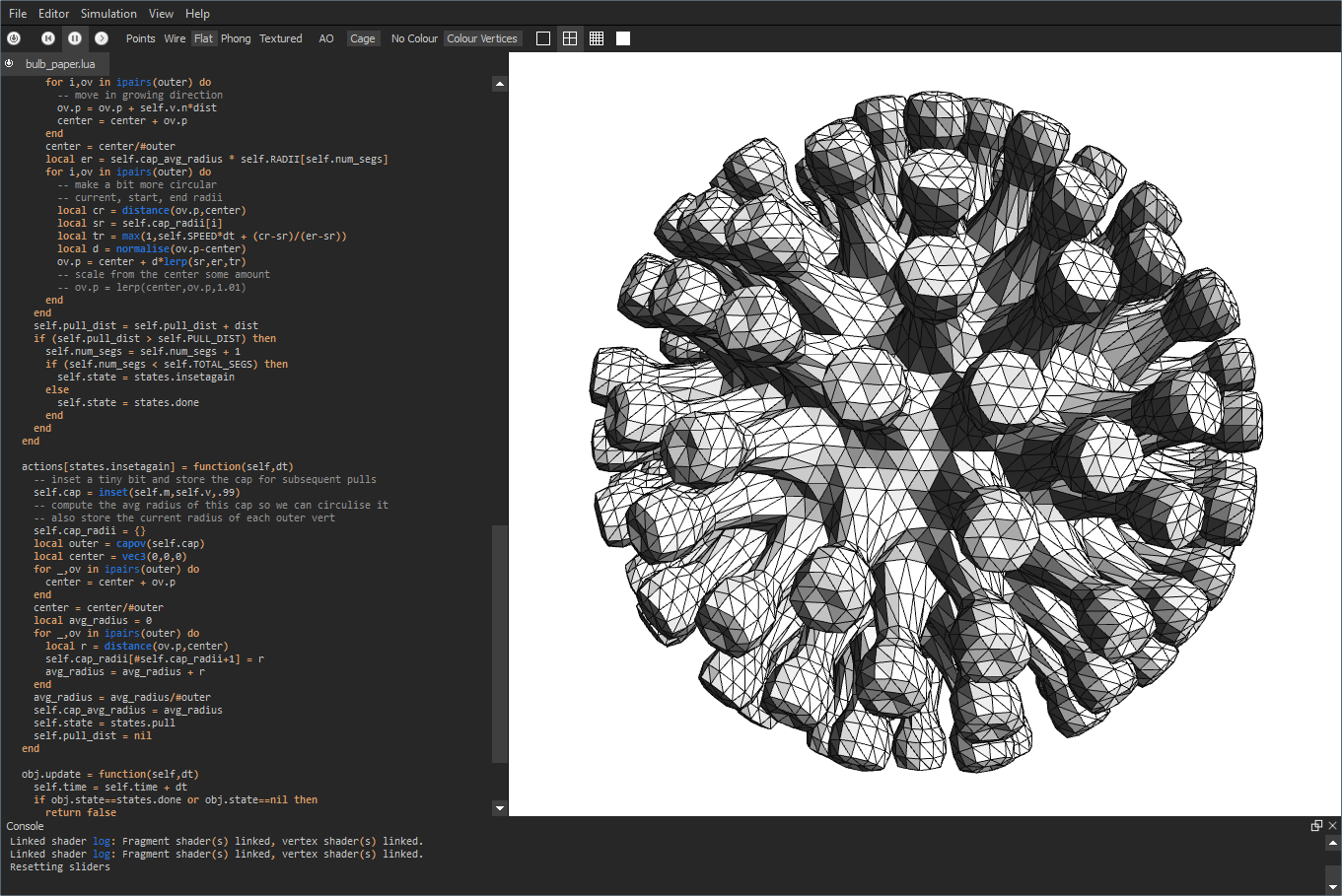}
  \caption{A screenshot of Fugu showing the code pane with syntax colouring (left), 3D interactive view of the model the script generates (right) and the console window for syntax and runtime error reporting (bottom).}\label{fig:screenshot}
\end{figure}

Fugu's interactive, code-oriented interface consists primarily of a code pane (Figure \ref{fig:screenshot}, left) and a 3D view (Figure \ref{fig:screenshot}, right). The design will be familiar to anyone who has used other popular creative-coding systems such as \emph{Processing}. Users can edit a script, press PLAY and see its effect immediately. They can also interact with the generated model while the simulation is running.

The code pane supports standard code editing functionality including syntax highlighting, line numbering, and multi-file editing with tabs. The 3D view uses a trackball-style manipulation of the scene viewport, and viewing modes range from wireframe to ambient occlusion shaded and textured modes.  Additionally the user can select smoothly subdivided versions of the geometry (using Butterfly subdivision \cite{Dyn1990}). A console window at the bottom of the screen presents script syntax and run-time errors.

\subsection{Script Format}\label{ss:fugu:script}

A Fugu script is a single file of Lua code written in the form of a Lua module. When a script is loaded, Fugu looks for two special functions in the module: \func{setup()} and \func{update()}.

The setup function is run once at the start of a simulation to specify initial conditions and create any initial meshes. For example, the setup function of Listing \ref{code:intro} creates a new spherical mesh, stores it in a module-scoped variable \emph{m}, then adds the mesh to the Fugu \emph{universe} (see Section \ref{ss:fugu:functionality}).

The update function is called repeatedly and at regular intervals from the time the user presses the PLAY button. Updating continues until the PAUSE button is pressed. The update function receives a parameter specifying the time elapsed since the last update, \emph{dt}. In Listing \ref{code:intro}, the update function iterates over all the vertices in the mesh \emph{m} and modifies their positions every time step.

\begin{lstlisting}[float,caption={An example Fugu script. The setup function creates a spherical mesh. During each update, all the vertices of this mesh are perturbed by a sine function that varies over time and local vertex position.},label=code:intro]
module(...,package.seeall)
local m
function setup()
    m = sphere()
    fgu:add(meshnode(m))
end
function update(dt)
    for _,v in ipairs(vertexlist(m)) do
        v.p = v.p + vec3(0,0.01*sin(v.p.x+fgu.t),0)
    end
end
\end{lstlisting} \label{code:intro}



\subsection{Geometric Modelling Functionality}\label{ss:fugu:functionality}

In addition to the standard Lua library, Fugu provides access to a range of functions and datatypes that support 3D mesh modelling. Fugu's primary object is the 3D triangular mesh, composed of vertices, edges, and triangles. Fugu simplifies working with mesh geometry by facilitating mesh creation and manipulation through an extensive set of functions and datatypes. A summary, classified according to type, is given below.

The \func{universe} is a simple scene graph for organising multiple mesh objects and their display. The userdata universe object \func{fgu}, contains all the objects of the scene and provides two member functions \func{add(n:node)} and \func{make\_child\_of(parent:node, child:node)} which allow a script to add a node to the universe and to anchor a node's position to another node respectively. There are two types of nodes: \emph{abstract nodes} which can be used as invisible anchors (e.g., as a pivot for an object), and \emph{mesh nodes}, which transform meshes in the scene. Mesh objects must be wrapped in a \func{meshnode} datatype before they are added to the universe (Listing \ref{code:intro}).

Affine \textbf{transformations} (stored as homogeneous matrices) transform nodes in the scene graph, or can be applied directly to a mesh to permanently transform its geometry. Fugu provides a shorthand for the standard geometric transformation matrices: \func{T(t:vec3)} translates a point by the given vector; \func{S(s:vec3)} scales by the vector \code{s}; \func{R(rad:double, axis:vec3)} rotates a point a number of radians around the specified axis; and, \func{Rv(a:vec3,b:vec3)} rotates vector \code{a} to align with vector \code{b}.

A number of \textbf{mesh creation} functions are provided. Standard primitives are created using \func{cube()}, \func{sphere()}, \func{icosahedron()}, etc. The \func{iso(r:int, f:function)} function generates an isosurface mesh by sampling the function \code{f} using the marching cubes algorithm within a 2x2x2 cube with resolution \code{r}. For instance, a sphere could be generated with the following statement:
\begin{lstlisting}
iso(16, function(x,y,z)
    return distance(vec3(x,y,z),vec3(0,0,0)) - 1
    end)
\end{lstlisting}
This example also illustrates the use of Lua's anonymous functions. Meshes can be loaded from files (in the popular .obj format), useful for applying functions to pre-built geometry (such as the Stanford Bunny in Figure \ref{fig:intro} (b)).

The creation of \textbf{generalised cylinders}---geometric surfaces defined by connecting a series of cross-section curves distributed along a \emph{carrier curve} \cite{Agin1972}---is also supported. A triangular mesh approximating the surface is constructed from the generalised cylinder, using a novel \emph{curve-morphing} technique \cite{Wetter2011}.

Cylinders are defined using a Logo-inspired turtle interface, in which a co-ordinate reference frame, the \func{turtle}, is created and transformed through space, tracing out cross-sections and carrier curves \cite{McCormack2004}. The turtle's member functions change its position and orientation, these include \func{move(d:double)}, \func{roll(a:double)}, \func{pitch(a:double)} and \func{yaw(a:double)}. Cross-section and carrier curves are modelled as piecewise cubic B\'{e}zier curves, the control points of which are added by the turtle as it moves through space, using \func{add\_point()}. After defining the cylinder, the member function \func{mesh()} creates and returns a triangular mesh version of the cylinder, on which additional mesh-based operations can then be performed. The example in Figure \ref{fig:complex} illustrates a complex organic form created in Fugu with generalised cylinders.

\textbf{Mesh Implementation}: Rather than design yet another mesh representation and manipulation library of our own, we used \emph{VCGLib} to provide the mesh functionality required in Fugu. VCGLib is a comprehensive C++ template library that provides a flexible framework for creating and manipulating triangular meshes (\url{http://vcg.sourceforge.net}). Using VCGLib permitted rapid development, allowing us to focus on designing new geometric operators and manipulators, leaving VCGLib to take care of mesh representation and geometric integrity. Additionally, VCGLib provides many useful operations, such as \emph{Loop subdivision}, and has been extensively used and tested in the popular program \emph{MeshLab} (\url{http://meshlab.sourceforge.net/}). Two alternative libraries, \emph{CGAL} (\url{http://www.cgal.org}) and \emph{OpenMesh} (\url{http://www.openmesh.org}), provide similar functionality and were considered for our application. CGAL is oriented towards meshes with a fixed topology and is more focused on guaranteeing correctness of algorithms with its precise kernels, rather than being an efficient real-time format. OpenMesh is a polygonal mesh library based around a half-edge data structure and associated operations. In retrospect OpenMesh may have been a better choice due to its simpler API and more liberal license than VCGLib (LGPL vs GPL).

The lowest-level of mesh manipulation occurs on \emph{vertices} and \emph{faces}. Following the design inherited from our mesh representation library, \emph{VCGLib} (see sidebar \textbf{Mesh Implementation}), edges are manipulated implicitly by modifying vertices and faces. The \func{vertexlist(m:mesh)} function provides access to a mesh's vertices as a Lua array. A \func{vertex} is a user-data object, and has a number of \emph{attributes} including a position, \code{p}, colour, \code{c}, and normal, \code{n}. Listing \ref{code:intro} illustrates one way a vertex position can be modified in a script. Should a script retain a reference to a vertex for some time, there is the possibility the reference may become invalid if the vertex is deleted by another part of the script. To protect against such issues, vertices (and faces) are modelled as \emph{proxies} (see sidebar \textbf{Proxies}). Faces have a similar set of functions, for example \func{facelist(m:mesh)} returns a list of faces in the mesh, \func{face.n} returns the normal of a face, and \func{face:v(i)} returns the i'th vertex of the face.

\begin{figure}[htbp]
  \centering
  \includegraphics[width=.5\textwidth]{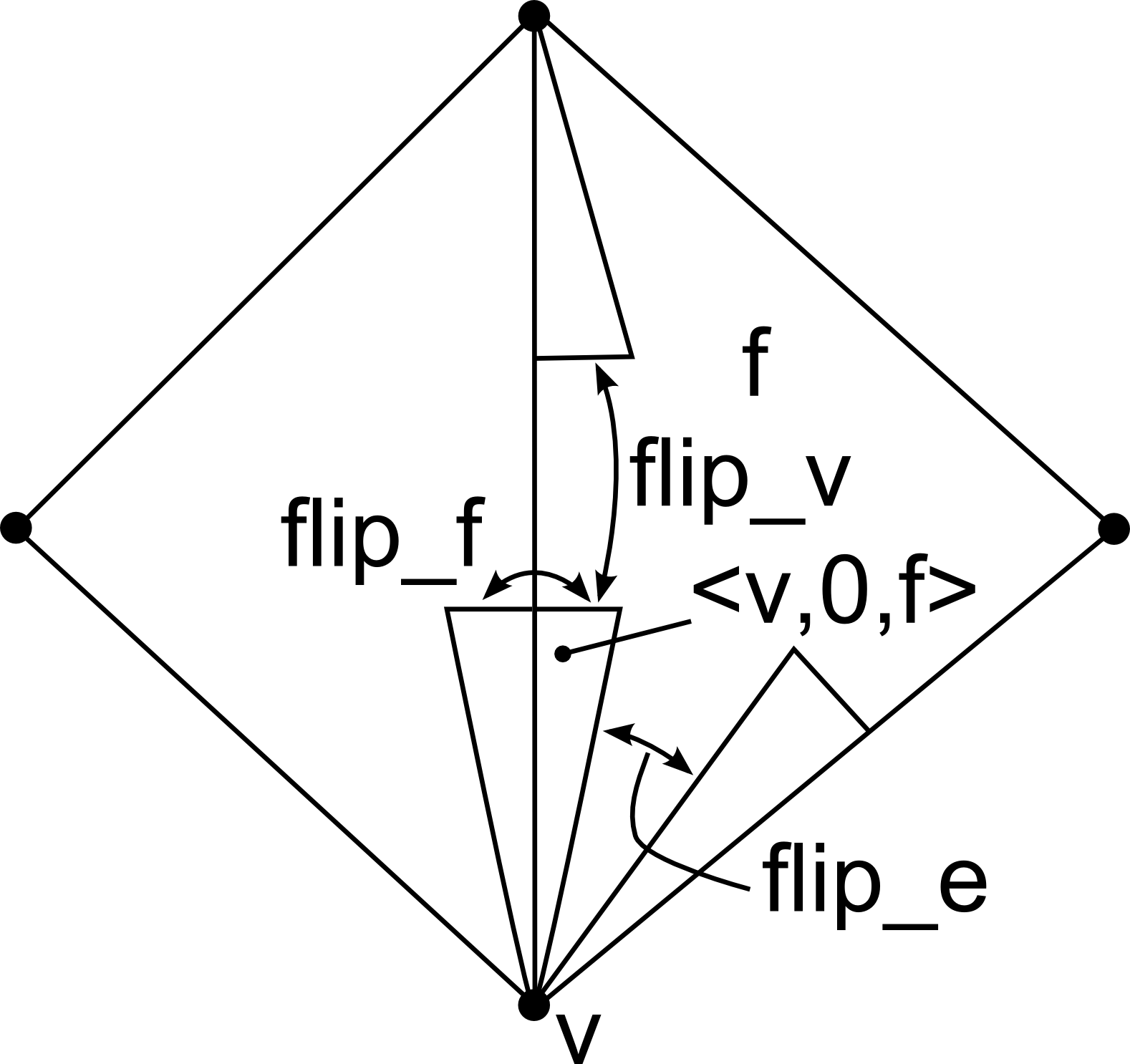}
  \caption{Using a \func{pos} to navigate a mesh. This figure shows two faces of a mesh and four pos'es, drawn as triangles connecting the vertex, edge and face referenced by each pos. From the pos \code{<v,0,f>} (indicated in the figure), we can move to the vertex above it with \func{flip\_v}, the opposing edge within the face with \func{flip\_e}, or to the adjacent face with \func{flip\_f}. If the mesh is fully connected we can reach any position on the mesh using a sequence of these operations.}\label{fig:pos}
\end{figure}

When modifying mesh geometry or vertex positions, normals may need to be recalculated to ensure correct shading and display. This is performed automatically between update calls or by calling \func{mesh:sync()} explicitly.

Fugu provides a simple mechanism for navigating around a mesh: the \func{pos} object, which models a cell-tuple \cite{Brisson1989}. A pos is a \code{<vertex,edge,face>} tuple, where vertex and face are the data structures introduced above, and edge is either 0 or 1 (referencing one of two possible edges). A pos uniquely identifies a position on a mesh and provides a set of member functions (\func{flip\_v()}, \func{flip\_e()}, and \func{flip\_f()}) for navigating it over the mesh (see Figure \ref{fig:pos}). Many of the functions in Fugu return a pos or a list of pos'es. Given a pos, \code{p}, the vertex or face it points to can be retrieved as \code{p.v} and \code{p.f}.

\textbf{Proxies}: Fugu is designed to allow multiple scripts to run simultaneously so that, for example, a physics script could simulate the effects of soft-body dynamics on a developing form, while a sprouting script could be sprouting florets from that form's surface. For simplicity we assume that scripts are not running truly concurrently, but rather that they have \code{update()} functions that are called sequentially. However, what happens when a script stores a reference to a vertex, and another script deletes that vertex? To safeguard against accessing an invalid object, vertices and faces are stored by proxy. This adds a safe layer of indirection, and offers a \func{valid()} function to check if the element targeted for access still exists. VCGLib also shuffles vertices and faces around in memory as elements are deleted, for space efficiency -- another reason that necessitates this safety mechanism. The vertex and face proxies implemented in Fugu also ensure they are updated to point to valid memory locations.

A number of \textbf{mesh operations} for assisting with modelling are provided. \emph{Accessor} functions return sets of elements: \func{loopv(v:vertex)}, for example, returns an ordered list of vertices that loop around \code{v}, and \func{nearbyv(v:vertex, n:int)} returns a list of vertices that are n edges or less away from v. Other functions, such as \func{mesh:smooth\_subdivide(levels:int)} operate on an entire mesh. VCGLib provides a large number of these functions, several of which we expose as Lua functions.

Local operations, such as \func{inset(v:vertex,s:double) }(demonstrated in Section \ref{ss:example:extrusion}), insets the faces surrounding a vertex and scales them by a given amount, \code{s}. \func{extrude(v:vertex, d:vec3, m:double)} extrudes the faces surrounding a vertex \code{v} in direction \code{d}, by magnitude \code{m}. Given a list of vertices, \code{vl}, and a plane defined with position \code{p} and normal \code{n}, \func{flattenvl(m:mesh, vl:list, p:vec3, n:vec3)} flattens all the vertices in the list to lie on that plane.

\textbf{Mathematics} functions include trigonometric functions, linear algebra, Perlin noise and a variety of interpolative functions. Datatypes for 3D vectors, homogeneous matrices, and quaternions are available as the userdata types \func{vec3}, \func{mat4}, and \func{quat} respectively. Lua's operator overloading features allow standard mathematical operators (addition, multiplication, etc.) to work on these new types (see Listing \ref{code:intro}).

A set of \textbf{geometric functions} are available to assist with performing geometric queries, collision tests, etc. For example, the function \func{distribute\_\ points\_sphere(n:int)} returns \code{n} equally distributed points on a sphere and \func{perp(v:vec3)} returns a normalised vector perpendicular to \code{v}.

A number of \textbf{utility} functions for manipulating Lua structures are provided, most of which are adopted from the \emph{Underscore.lua} library (\url{http://mirven.github.com/underscore.lua/}). These functions provide a simplified syntax for iterating over Lua tables and performing functional programming. For example, the \func{each(t:table, f:function)} utility function applies \code{f} to each member of \code{t}.

Higher-level functions are implemented directly in Lua, while lower-level, CPU intensive functions are implemented in C++. The support library, \emph{Luabind}, was invaluable in allowing easy binding between C++ datatypes and variables and Lua. Our goal is to eventually provide enough functionality so that most mesh operations can be coded purely on the Lua-side, allowing user-created libraries to be easily shared amongst the user community.

Having now described the basics of Fugu operation, along with its principal functions and datatypes, we will illustrate how these features can be used to model 3D form using a series of examples.

\section{Example Applications}\label{s:example}

In this section we describe two applications that highlight the different modelling features available within Fugu. The first example demonstrates a mesh operation for creating continuous animated extrusions. This compound extrusion operation (implemented as a Lua function that combines several lower-level operations) is used to generate a variety of tubular organic structures over existing meshes. The example also illustrates the general structure of a Fugu script and key mesh-level operations. The second example describes a timed L-system simulation, based on \emph{Calcispongiae} development. Unlike previous L-system modelling tools, the geometry generated by the L-system's development can be subject to further modification using Fugu's mesh operations.

\subsection{Extrusion}\label{ss:example:extrusion}

Extrusion is a fundamental operation in 3D mesh modelling that involves the translation of a group of triangles, typically along their normal, to generate a new form over the region that is swept out. A continuous extrusion can be used to animate the outward growth of a limb, spike, or thorn. This section illustrates how continuous extrusions are modelled and applied over arbitrary meshes in our system. By changing extrusion parameters a variety of different effects can be achieved.

The \func{extrude(v:vertex,dir:vec3,m:double)} function extrudes a set of faces adjacent to a specified vertex, along a supplied direction by amount \code{m}. To effectively model continuous growth, discontinuities in the surface geometry must be minimised as it undergoes extrusion. One method of achieving this is to use very small extrusion steps. However, in the model described below, we use the function, \func{inset}, which performs a zero-distance extrusion, followed by scaling of the end faces.

A continuous extrusion operator can be modelled with a \emph{move phase}, during which the vertices of the extrusion are translated continuously in a specified direction; and an \emph{inset phase}, where the faces at the end of the extrusion are inset, forming a new cap to translate. It is useful to consider growth as a single entity, so to model this in Lua we create a Lua object with an internal state and an \func{update} function that is responsible for performing the extrusion. This \func{update} function returns \code{false} when the extrusion is complete, and \code{true} otherwise. The internal logic is modelled using a state machine with three states: \code{move}, \code{inset}, and \code{done}. The script for this object is shown in Listing \ref{code:new_spike}.

\begin{lstlisting}[float,caption={This function creates a Lua object which, by repeatedly calling its update member function, will create an extrusion at the supplied vertex.},label=code:new_spike]
function new_spike(the_mesh,the_vertex)
    -- the possible states
    local states = {
        move = 1,
        inset = 2,
        done = 3
    }

    -- constants for the operation
    local SPEED = 4	
    local SEG_LENGTH = .1
    local NUM_SEGS = 5
    local SHRINK = .8
	
    -- create the object and its initial instance variables	
    local obj = {
        m=the_mesh,
        v=the_vertex,		
        n=the_vertex.n,	
        seg = 1,
        distance = 0,
        cap = nil,		
        state=states.move
    }	
	
	-- define the actions
	local actions = {}	
	actions[states.move] = function(self,dt) ... end
	actions[states.inset] = function(self,dt) ... end

	-- the update function executes the action based on its state
	obj.update = function(self,dt)
		if self.state==states.done then
			return false
		else
			actions[self.state](self,dt)
			return true
		end
	end	
	
	--  return the new object
	return obj
end
\end{lstlisting}

The inset state executes the \func{inset} function (Listing \ref{code:actions}) and then returns the machine to the move state. This function doesn't change \code{self.v} (the vertex currently being extruded), which always refers to the vertex at the end of the extrusion. Additionally, the \func{inset} function returns the \emph{cap}: a fan of pos'es (cell-tuples) iterating over the faces located at the active end of the extrusion. This is used in the \code{move} state to access the vertices surrounding \code{self.v}.

\begin{figure}
  \centering
 \includegraphics[width=\textwidth]{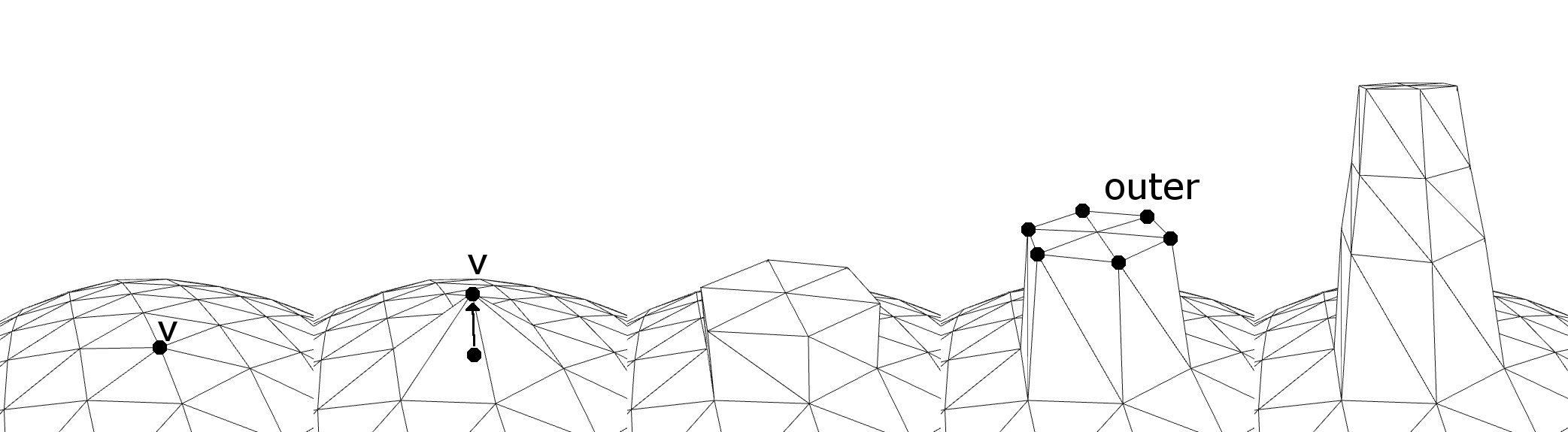}
  \caption{A sequence generated by applying the \code{new\_spike} operation in Listing \ref{code:new_spike} to vertex \code{v}. The sequence alternates between moving the vertex (and its neighbours) and insetting a new set of faces on the end of the protrusion. The variable \code{outer} in Listing \ref{code:actions} refers to the vertices at the end of the cap (as shown).
  }\label{fig:extrusionexample}
\end{figure}

While in the move state, the extrusion shifts \code{self.v} by a small amount in the normal direction, then moves the adjacent vertices extracted from \code{self.cap} using the function \func{capov(cap:list)}, which, for a given pos-fan, returns the outer vertices as a Lua array. The vertices are first moved in the extrusion direction, and then scaled from the center so that they shrink. The final step calls \func{flattenvl(m:mesh,vl:list,p:vec3,n:vec3)}, which flattens the end cap vertices in the list \code{vl}, so they sit on the plane defined by position \code{p} with normal \code{n}. If the vertices have been shifted more than \code{SEG\_LENGTH}, then the state is either changed to \code{inset} to generate a new segment, or to \code{done} if the required number of segments have been generated. Figure \ref{fig:extrusionexample} illustrates these stages in the extrusion process.

\begin{lstlisting}[float,caption={The two actions corresponding to the inset and move phases.},label=code:actions]
actions[states.inset] = function(self,dt)
	self.cap = inset(self.m,self.v,.99)
	self.state = states.move
	self.distance = 0
end

actions[states.move] = function(self,dt)
	local dist = SPEED*dt
	self.v.p = self.v.p + self.n*dist
	if (self.cap) then
		local outer = capov(self.cap)
		local center = vec3(0,0,0)
		for _,ov in ipairs(outer) do
			ov.p = ov.p + self.n*dist
			center = center + ov.p
		end
		center = center/#outer			
		for _,ov in ipairs(outer) do
			ov.p = center + (ov.p-center)*SHRINK
		end
		flattenvl(self.m,outer,self.v.p,self.n)
	end			
	self.distance = self.distance + dist			
	if (self.distance > SEG_LENGTH) then
		self.seg = self.seg + 1
		if (self.seg <= NUM_SEGS) then
			self.state = states.inset
		else
			self.state = states.done
		end
	end		
end
\end{lstlisting}

\begin{figure}
    \centering
    \subfigure[]{
    \includegraphics[width=.3\textwidth]{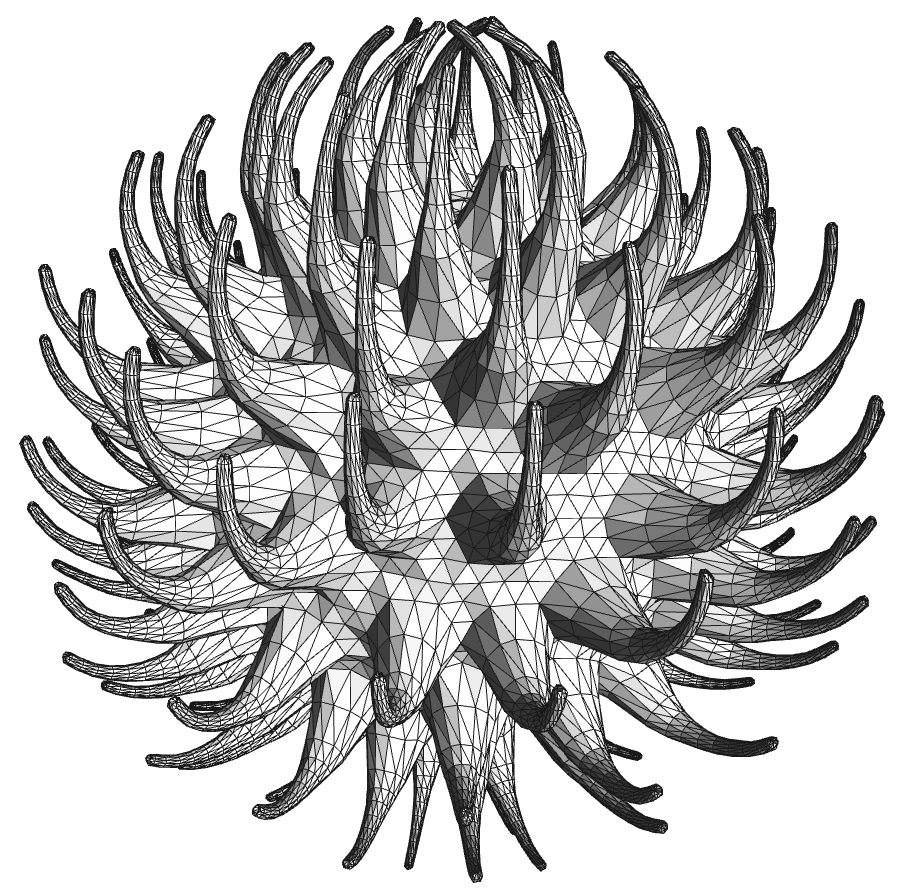}
    \label{fig:extrusions:a}
    }
    \subfigure[]{
    \includegraphics[width=.3\textwidth]{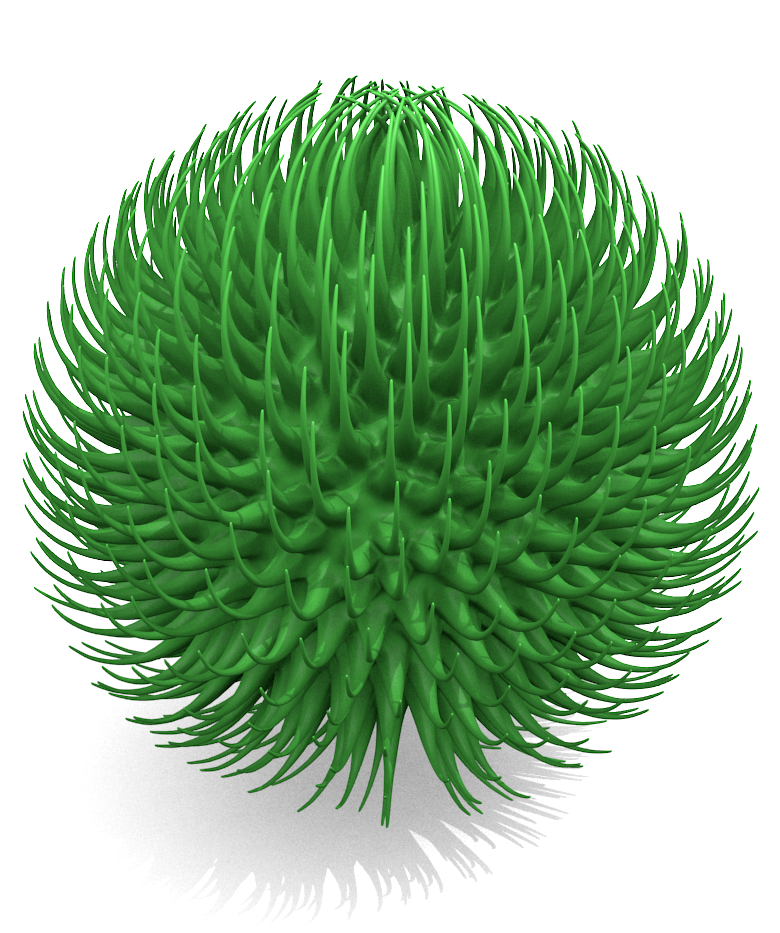}
    \label{fig:extrusions:b}
    }
    \subfigure[]{
    \includegraphics[width=.3\textwidth]{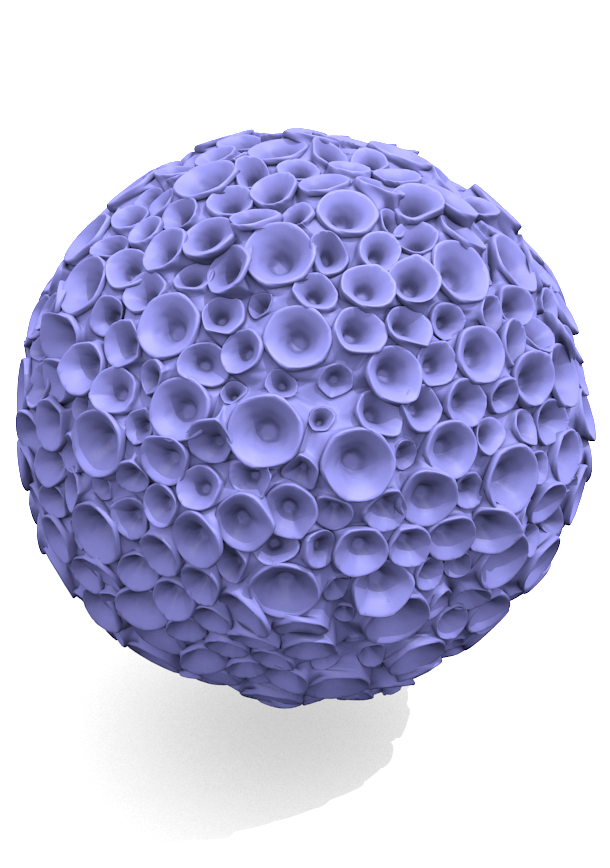}
    \label{fig:extrusions:c}
    }
    \label{fig:extrusions}
    \caption{(a, b) Forms created by growing tapered extrusions out of a sphere. The spikes curve to orient towards a point above the sphere. (c) The suckers on this object are generated by extruding outwards and then inwards. A collision routine ensures that the suckers only grow until they touch a neighbouring sucker.}
\end{figure}

Useful modifications to the continuous extrusion operation include rotating the end cap each segment, making the end cap more circular (to reduce the effect of starting conditions), and creating heterogeneous forms by modifying the extrusion object parameters based on vertex height (see Figure \ref{fig:intro} (a) and Figure \ref{fig:extrusions} (a,b)). By extruding outwards and then \emph{inwards}, we can create more interesting shapes, such as suckers (Figure \ref{fig:extrusions} (c)). This extrusion operation is general enough to be performed on any smooth manifold mesh. Figure \ref{fig:intro} (b) illustrates the operation on the Stanford Bunny, for example.

\subsection{Generalised Cylinders}\label{ss:example:lsystem}

In addition to mesh modification functions, Fugu contains a number of mesh generation methods, the most sophisticated of which is the \emph{generalised cylinder}. In this section we demonstrate how we can easily model timed, parametric L-systems in Lua and use them to drive Fugu's generalised cylinder routines, creating animated growth of organic forms. We then show how this geometry can be manipulated further, by combining the L-system model with the extrusion model introduced in the previous example.

\subsubsection{Modelling L-systems with Fugu}

L-systems, introduced by Lindenmayer \cite{Lindenmayer1968}, are string rewriting grammars commonly used to model the development of cellular structures, herbaceous plants and trees \cite{Prusinkiewicz1996}. \emph{Parametric L-systems} represent models as symbolic strings with associated scalar parameters, which develop from an initial symbol string (an \emph{axiom}) according to a set of rewrite or \emph{production rules}. To obtain geometry from an L-system, the produced string must be interpreted by a geometry building mechanism.

Lua tables are a convenient and efficient data structures for storing an L-system string and specification, but from a user perspective they are too verbose. Therefore, we allow a user to define an L-system using Lua strings (and arrays of strings), which are parsed using Lua's string matching library (see Listing \ref{code:lsys}).

\begin{lstlisting}[float,caption={Creating a parametric L-system in Lua.},label=code:lsys]
axiom = 'B(2) A(4,pi+1)'
rules = {
    'A(x,y) : y <= 3 -> B(x) A(x*2,x+y)',
    'B(x)   : x < 1 -> C(noise(x))',
    'B(x)   : x >= 1 -> B(x-1)'
}
lsys = new_lsystem(axiom,rules)
\end{lstlisting}

The function \func{new\_lsystem} creates an L-system object containing the current string (a table of symbols and a table of associated parameters) and the production rules. For example, the first L-system string contained in the object, \code{lsys}, in Listing \ref{code:lsys} would be represented as a table with contents:
\begin{lstlisting}
symbols = {'B','A'},
parameters = {{'2'},{'4','pi+1'}}
\end{lstlisting}
Production rules are also stored using Lua tables. The third rule in Listing \ref{code:lsys} would thus be a table containing the following:
\begin{lstlisting}
pred = 'B'
parameters = {'x'}
condition = 'x>=1'
succ = {'B'}
succ_par = {{'x-1'}}
\end{lstlisting}
The L-system object has a member function, \func{derive()}, which produces a derivation string by iterating through the current string and, for each symbol, checking if both the rule predecessor symbol matches and the production conditions are met. If this is the case, the symbol is replaced according to the matched rule.

Parameters and conditions are stored as strings so they can be evaluated dynamically by the Lua interpreter as production rules are matched and applied. This has the benefit of allowing any valid Lua expression (including Lua or Fugu functions) to be used in a parameter expression or condition; for example, the second rule in Listing \ref{code:lsys} illustrates the use of Fugu's \func{noise} function.

At this stage the L-system is purely symbolic, it is then interpreted to generate mesh geometry using the \emph{turtle interpretation} provided by Fugu's generalised cylinder functionality (Section \ref{ss:fugu:functionality}).

\subsubsection{Timed Development}

Parametric L-systems provide a discrete-time model of development, making continuous temporal development difficult. To overcome this limitation, \emph{timed parametric L-systems} were introduced \cite{Prusinkiewicz1996}. Timed L-system symbols are assigned an \emph{age} -- a continuous variable representing the time the symbol has been active in the derivation string. This variable also determines when a production rule should be applied to its associated  symbol. A symbol's age can also be used to drive other animation parameters, such as a primitive's scale or length \cite[Chapter 6]{Prusinkiewicz1996}.

Our Lua-based implementation of parametric L-systems can be easily extended to include timed symbols. The L-system object stores an additional table with the age of each developing symbol in the produced string. Additionally, production rules may include a \emph{terminal} age, and a \emph{minimum age} of the predecessor in order for the rule to be applied.

Timed, parametric L-systems can be used to generate animated meshes. A Fugu script first defines an L-system object in its \func{setup} function. The L-system's member function \func{derive(dt:double)} is then called within the script's \func{update} function. The \func{derive} function updates each active symbol's age by \code{dt}, and then applies the appropriate production rules as necessary. Once the state of the L-system has been updated, it is re-interpreted to generate a new mesh.

\begin{figure}
    \centering
    \includegraphics[width=\textwidth]{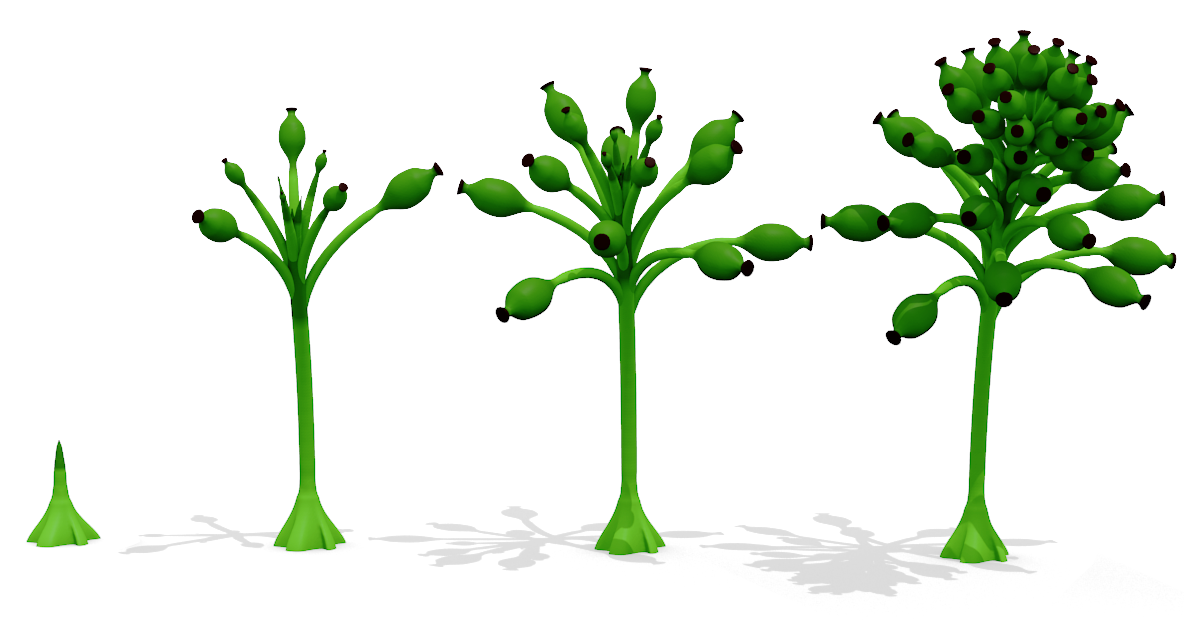}
  \caption{A growth sequence of a timed, parametric L-system built in Fugu.}\label{fig:calci}
\end{figure}

Using this method, we designed a timed, parametric L-system in Fugu to produce a form inspired by \emph{Calcispongiae} \cite{Haeckel1904}. The result is an animated sequence of geometric models with smooth and fluid continuous development. Figure \ref{fig:calci} shows a sequence of still images from this development. The appendix contains details of the L-system used.

\begin{figure}
    \centering
    \includegraphics[width=\textwidth]{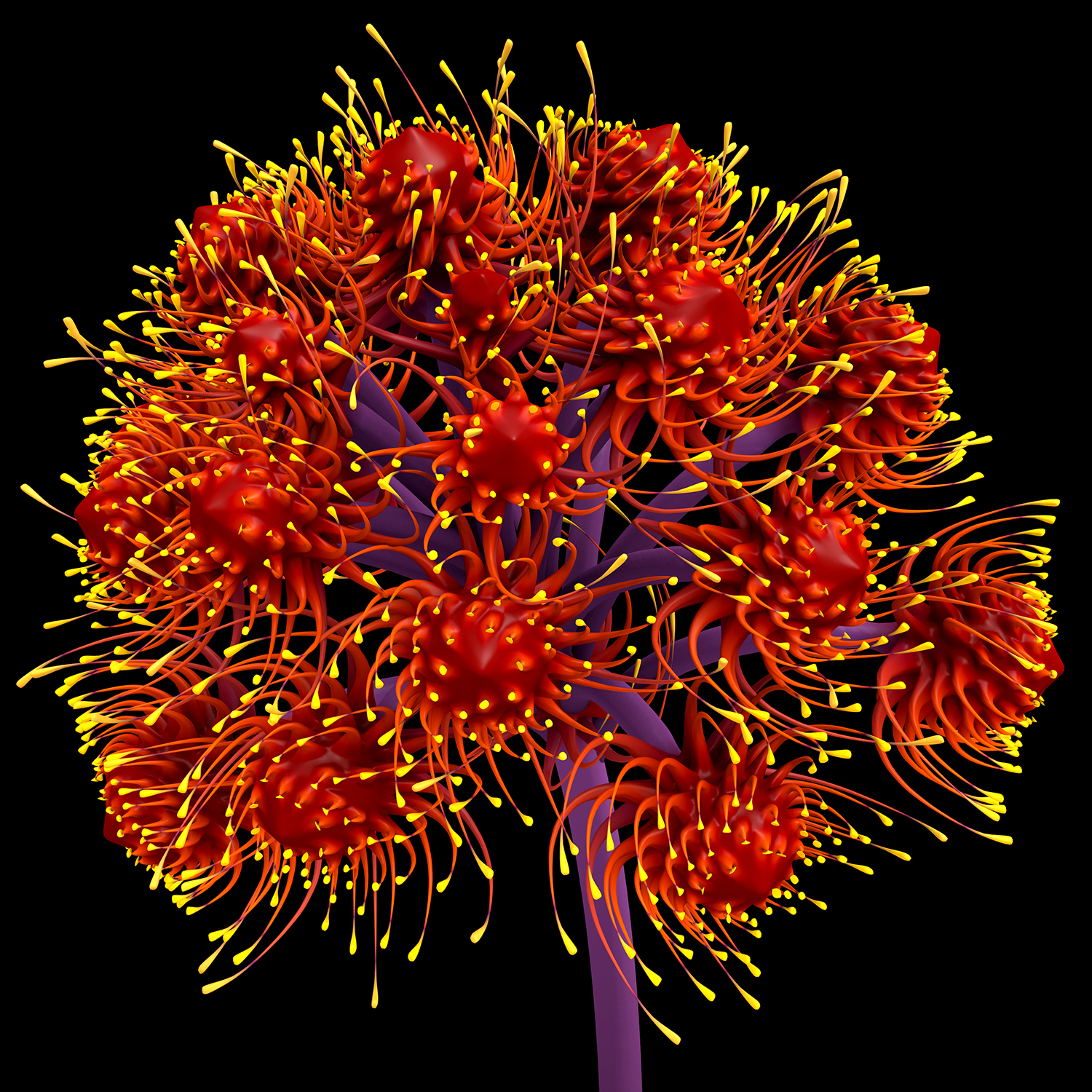}
    \caption{A flower-like form built with a combination of L-systems, generalised cylinders, and mesh extrusions. This hybrid model combines the models used to generate Figure \ref{fig:calci} and Figure \ref{fig:intro} (a).}\label{fig:complex}
\end{figure}

The L-system generates a triangular mesh, which can then have other mesh operations applied to it. Trivially we can just return the plain mesh and operate on it, but more interesting behaviour can be modelled if information is shared between the L-system model and Lua. For the model shown in Figure \ref{fig:complex}, we used texture coordinates assigned to the generalised cylinders to control the placement and properties of extrusions grown on the mesh.

This idea could be extended to allow general tagging or identification of mesh parts. For example, the bulbs in Figure \ref{fig:calci} are created by a substring in an L-system, we could delimit that substring with special symbols that cause the bulb part of the mesh to be tagged (either with a per-vertex attribute or by returning a list of faces associated with tags). The bulb could be tagged with any information, including a symbols age or parameter, with which a mesh operation could act. For instance, the bulb's age could affect a hypothetical \emph{wrinkle} operation, causing the older buds to wrinkle more than the younger ones.

As these examples show, the procedural flexibility of scripts, combined with powerful mesh manipulation and development functions makes the specification of complex models relatively easy.

\section{Conclusions}\label{s:conclusions}

The benefits of procedural design are well known, but the accessibility of advanced procedural modelling  tools to designers and architects has, until recently, been limited. With new code-based creative tools such as Fugu, this goal becomes increasingly viable. However, there are many additional features that could extend the capabilities of our system to increase its versatility.

The most useful enhancement would be multiple, complementary model representations. For example: a geometric mesh, an armature for skeletal animation and construction, and a physical simulation system. Ideally, these different representations can be flexibly connected together. This could, for example, allow a user to procedurally generate physical structures that could be built using automated building techniques, such as multi-axis robots. The armatures could in turn define geometric primitives used in a physical simulation, preventing the developing structure from intersecting and allowing it to move in response to physical changes.

Morphogen simulation (based on biological patterning) is often used to generate a variety of self-organising natural structures and patterns, and has been widely used in modelling development \cite{Porter2010}. Allowing user-defined attributes for the vertices and faces would facilitate this feature in Fugu, and permit greater flexibility in modelling development. For instance, a user could specify that their model requires a new, per-vertex variable \emph{morphogen}. This could also extend to per-face attributes, for instance to define material properties for rendering. Simulation of morphogen diffusion could be implemented as additional Lua functions, allowing users to easily modify developmental behaviour or implement unusual effects within scripts.

The functionality in Fugu has been oriented towards modelling of specific target forms, such as abstract organic shapes. A number of additional operations and features would expand Fugu's modelling possibilities, while still operating within the script-oriented paradigm. Low-level mesh operations, such as adding and removing vertices and faces, welding vertices, and flipping edges, are necessary for more complex operations to be specified at the Lua level. Here a formal algebra could be employed, such as \emph{Vertex-Vertex} systems \cite{Smith2006}. Higher-level mesh operations such as bevelling, re-triangulation, and edge ring selection could easily be added based on functionality already present in VCGLib.

There are also some technical challenges that need to be overcome to support the real-time creation of more complex models. The smooth-subdivision visualisation used in the current system is performed in a single-thread on the CPU, and thus impacts significantly on overall performance when active (viewing more than 2 levels of subdivision is not currently practical for interactive performance on standard desktop computers). GPU-based smooth subdivision using tessellation shaders found in recent hardware would dramatically improve visualisation performance (see e.g.\ \url{www.opengl.org/registry/specs/ARB/tessellation_shader.txt}).

Lua co-routines offer a parallelism mechanism that could be used to simplify the specification of models \cite{Moura04}. For instance, an operation that performs a sequence of actions, such as \emph{grow stalk}, \emph{grow flower}, \emph{detach petals}, each requiring a second of simulation time to complete, could be implemented like this:
\begin{lstlisting}
obj.update = update(self,dt)
    for _,f in ipairs{grow_stalk,grow_flower,detach_petals} do
		local time = 0
		while (time < 1) do
			time = time + dt
			f(self,dt)
			coroutine.yield(true)
		end
	end
return false
end
\end{lstlisting}
Where \func{grow\_stalk}, etc., are member functions of the object. This is more concise and arguably a lot more readable than manually maintaining a state and state machine as is done in Listing \ref{code:new_spike}.

The core of Fugu has been designed to be independent of any visual representation, making it easy to incorporate it into a real-time game engine, and could take advantage of its visualization features  (e.g.\ shadows, post-processing, shaders). Further development would be required to ensure real-time frame rates however.

The novelty of our system derives from its combination of a rich set of geometric mesh operations with a fast and flexible scripting language. As similar systems have demonstrated in more general ``creative coding'' environments, the use of relatively simple function libraries combined with an initialisation and per-frame step programming model makes it easy for artists and designers to rapidly create animated sequences. Fugu brings this ease of use and experimental flexibility to 3D animated mesh modelling. We hope that as its development continues, Fugu will be able to tackle even more complex modelling tasks in the design of procedurally generated form for design and architecture applications.

You can download a copy of Fugu, including source code, from \url{www.csse.monash.edu.au/cema/fugu}.

\section{Acknowledgements}

This research was supported by an Australian Research Council Discovery Projects grant DP1094064.

\section{Appendix}

The timed, parametric L-system used to generate the \emph{Calcispongiae} shown in Figure \ref{fig:calci} is shown below. A symbol $X$ with parameters $p_1 \ldots p_n$ and age $\alpha$ is denoted by $(X(p_1,\ldots,p_n),\alpha)$. Non-timed symbols are also allowed, in which case the outer parentheses and age are omitted. The turtle commands for each symbol are specified in the table.

The L-system consists of three main components, modelled with symbols $A$, $B$ and $C$. Symbol $A$ generates the first section of the stalk before any branching has occurred, symbol $B$ is responsible for placing the branches in a helical pattern based on the golden angle, and symbol $C$ is responsible for generating each branch and its terminal bulbs.

\noindent \begin{tabular*}{\textwidth}{@{\extracolsep{\fill}}ll}
  \hline
\multicolumn{2}{l}{$\mbox{\textbf{axiom}}: \wedge(-.5\pi) G\#(1) (Gsc(3),0) Gs G\#(0) (f(6),0) (A(2,1,10,0.02),0) Ge $} \\
\multicolumn{2}{l}{\textbf{production rules}} \\
$(A(l,w,n,b),1) : n>0 $&$\rightarrow (S(l,b,w),0) (A(l,w,{n-1},b),0)$ \\
$(A(l,w,n,b),1) : n=0 $&$\rightarrow (B(2,0.7,25,0.09),0)$ \\
$(B(l,w,n,b),1) : n>0 $&$\rightarrow [\backslash(2.39996n) (Gsc(w),0) Gs (f(l),0) (C(l,w,10,b),0) Ge]$\\
&$\ldots (S(1,1,1),0) (B(l,w,n-1,0.9b),0)$ \\
$(C(l,w,n,b),1) : n>0 $&$\rightarrow (S(l,b,w),0) (C(l,w,n-1,b),0)$ \\
$(C(l,w,n,b),1) : n=0 $&$\rightarrow (S(1.5l,0,1.1l),0) (S(1.5l,0,l),0)$\\
&$\ldots (S(l,0,0.3l),0) (S(.5l,0,0.6l),0)$ \\
$(S(l,b,w),30)$&$\rightarrow S(l,b,w)$ \\
$(Gsc(x),10) $&$\rightarrow Gsc(x)$ \\
$(f(x),15) $&$\rightarrow f(x)$ \\

\multicolumn{2}{l}{\textbf{symbol interpretations}} \\
\multicolumn{2}{l}{$S(l,b,w)$: Add a segment to the generalised cylinder of length $l$}\\
\multicolumn{2}{l}{pitched by angle $b$, and ending with scale $w$}\\
\multicolumn{2}{l}{$G\#(n)$: Set the cross section}\\
\multicolumn{2}{l}{$Gsc(x)$: Set the scale for the next cylinder control point}\\
\multicolumn{2}{l}{$Gs$: Begin a generalised cylinder}\\
\multicolumn{2}{l}{$Ge$: End a generalised cylinder}\\
\multicolumn{2}{l}{$\backslash(x)$: Roll by $x$ radians}\\
\multicolumn{2}{l}{$f(x)$: Move forward $x$ units}\\
\multicolumn{2}{l}{$\wedge(x):$ pitch by $x$ radians}\\
\hline
\end{tabular*}

\bibliographystyle{plainnat}
\bibliography{references}

\end{document}